\documentclass[10pt,letterpaper]{article}

\usepackage[top=0.85in,left=2.75in,footskip=0.75in]{geometry}
\usepackage{graphicx,dcolumn,bm,fleqn,epic,eepic,float,ulem}
\usepackage{amssymb,amsmath,multirow,rotate,color,rotating}%,caption}
\usepackage[utf8x]{inputenc}
\definecolor{red}{rgb}{1,0,0}
\definecolor{green}{rgb}{0,1,0}
\definecolor{blue}{rgb}{0,0,1}

\begin{document}
%%%%%%%%%%%%%%%%%%%%%%%%%%%%%%%%%%%%%%%%%%%%%%%%%%%%%%%%%%%%%%%%%%%%%%%

\vspace*{0.2in}

\begin{flushleft}
% Please use "sentence case" for title and headings (capitalize only the first word in a title (or heading), the first word in a subtitle (or subheading), and any proper nouns).
{\Large
  \textbf\newline{Modeling specific action potentials in the human atria based on a minimal reaction-diffusion model}
%  to case-specific atrial physiology data}
}
\newline
\\
Yvonne Richter\textsuperscript{\ddag},
Pedro G.~Lind\textsuperscript{\ddag*},
Philipp Maass\textsuperscript{\ddag}
\\
\bigskip
 Fachbereich Physik, Universit\"at Osnabr\"uck, 
 Barbarastra\ss e 7, 49076 Osnabr\"uck, Germany
\\
\bigskip
\ddag
These authors contributed to this work in the following way: YR performed the simulations; PL wrote the paper, made the comparative analysis and derived the analytical results; PM revised the text and coordinated the research activities.
% Additional Equal Contribution Note
% Also use this double-dagger symbol for special authorship notes, such as senior authorship.
%\ddag These authors also contributed equally to this work.
%\textcurrency Current Address: Dept/Program/Center, Institution Name, City, State, Country % change symbol to "\textcurrency a" if more than one current address note
% \textcurrency b Insert second current address 
% \textcurrency c Insert third current address
% Group/Consortium Author Note
%\textpilcrow Membership list can be found in the Acknowledgments section.

% Use the asterisk to denote corresponding authorship and provide email address in note below.
* pelind@uos.de

\end{flushleft}

%%%%%%ABSTRACT
\section*{Abstract}
We present an effective method to model empirical action potentials of
specific patients in the human atria based on the minimal model of
Bueno-Orovio, Cherry and Fenton adapted to atrial electrophysiology.
In this model, three ionic are currents introduced, where each of it
is governed by a characteristic time scale.
By applying a nonlinear optimization procedure, a best combination of 
the respective  time scales is determined, which allows one
to reproduce specific action potentials with
a given amplitude, width and shape.
Possible applications for supporting clinical diagnosis are pointed out.

\section{Introduction}

Detailed reaction-diffusion models to describe human atrial electrophysiology were first developed in the late 1990s 
\cite{Courtemanche/etal:1998, Nygren/etal:1998, Luo/Rudy:1994, Lindblad/etal:1996} 
and are further developed until now.
Important steps forward have been made to
include specific ionic
currents~\cite{Courtemanche/etal:1999, Zhang/etal:2005, 
Maleckar/etal:2008, Tsujimae/etal:2008, Cherry/etal:2008, 
Koivumaeki/etal:2011}, which in particular allow one
to investigate specific effects of pharmaceuticals
in treatments of atrial fibrillation and other heart failures.
Complementary to these detailed models,
Bueno-Orovio, Cherry and Fenton  introduced in 2008 a minimal
reaction-diffusion model (BOCF model) for action potentials (AP) in
ventricular electrophysiology,
where the large number of ionic currents through cell membranes is reduced to 
three 
%most relevant 
net currents \cite{Bueno-Orovio/etal:2008}.
This model has four state variables, one describing the transmembrane voltage 
(TMV), and the other three describing the gating of ionic currents.
The TMV, as in detailed reaction models,
satisfies a partial differential equation of diffusion type with the 
currents acting as source terms, and the time evolution of the gating variables
is described by three ordinary differential equations coupled to the TMV.
By fitting the action potential duration (APD), the effective
refractory period and the conduction velocity
to the detailed model of Courtemanche, Ramirez and Nattel 
\cite{Courtemanche/etal:1998} (CRN model), 
the BOCF model was recently adapted
to atrial electrophysiology (BOCF model) \cite{Lenk/etal:2015}.
%%%%%%%%%%%%%%%%%%%%%%%%%%%%%%%%%%%%%%%%%%%%%%
\begin{figure}[!t]
\centering
\includegraphics[width=0.7\textwidth]{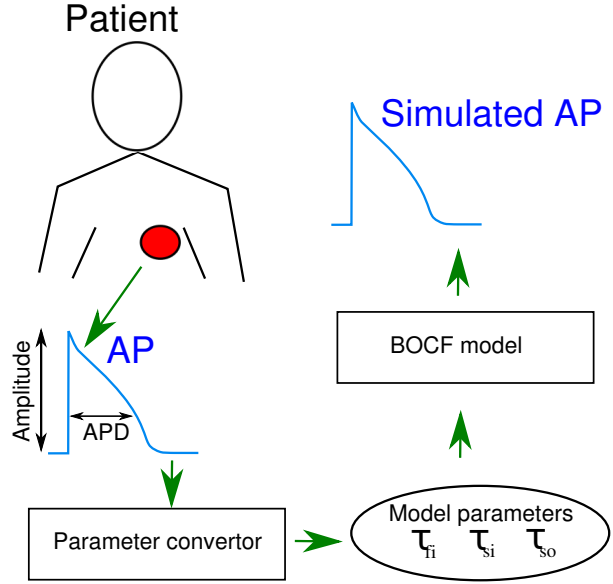}
\caption{\protect 
  \label{fig01}
  Schematic illustration of the optimized adjustment of
  the BOCF model by a parameter converter that determines
  the set of parameter values ($\tau_{\rm fi}$, $\tau_{\rm si}$,
  and $\tau_{\rm so1}$) giving a best match with the 
  amplitude and duration of the action 
  potential for a specific patient.}
\end{figure}
%%%%%%%%%%%%%%%%%%%%%%%%%%%%%%%%%%%%%%%%%%%%%%%%%%%%%%%%%%%%%%%

\begin{figure}[!t]
\centering
\includegraphics[width=0.6\textwidth]{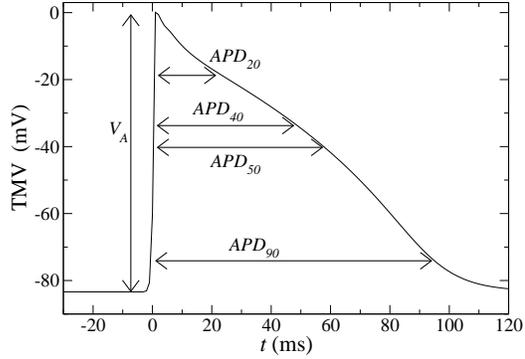}
\caption{\protect 
  Illustration of an action potential with amplitude $\hbox{APA}$ and
  four AP durations at 90$\%$, 50$\%$, 40$\%$ and 20$\%$ of the
  total amplitude.
  These five values are used to determine three characteristic
  time scales of the BOCF model (see text).}
\label{fig02}
\end{figure}

In this work we develop a method to model specific AP based on the BOCF model
as it is aimed in the clinical context in connection with improved and
extended possibilities of diagnosis~\cite{weber:2011}.
Compared to the detailed models, the BOCF model
has the advantage that it is better amenable to some analytical treatment.
This allows us to identify a small set of relevant model parameters
for capturing the main features of a specific AP.
Our methodology is sketched in Fig.~\ref{fig01} and can be summarized
as follows.
We start by labeling each given AP with its amplitude $\hbox{APA}$ and 
with four APD, namely at 90\%, 50\%, 40\% and 20\% repolarization, 
denoted as APD$_{90}$, APD$_{50}$, APD$_{40}$, and APD$_{20}$
respectively.
These APD$_n$ ($n=20$, 40, 50, 90) together with the amplitude $\hbox{APA}$ 
are suitable to catch a typical shape of a specific AP, see Fig.~\ref{fig02}.

The APD$_n$ taken for a specific
patient are given to a parameter convertor that retrieves
specific parameter values of the BOCF model. 
As relevant parameters, we adjust three time scales governing the
closing and opening of the ionic channels.
The parameter convertor consists of an optimization algorithm
that searches for the best set of parameter values consistent
with the measured AP properties.
%%%%%%%%%%%%%%%%%%%%%%%%%%%%%%%%%%%%%%%%%%%%%%%%%%%%%%%%%%%%%%%%%%%%%%%%
%%%%%%%%%%%%%%%%%%%%%%%%%%%%%%%%%%%%%%%%%%%%%%%%%%%%%%%%%%%%%%%%%%%%%%%%
%%%%%%%%%%%%%%%%%%%%%%%%%%%%%%%%%%%%%%%%%%%%%%%%%%%%%%%%%%%%%%%%%%%%%%
\begin{figure}[htb]
\centering
\includegraphics[width=0.9\textwidth]{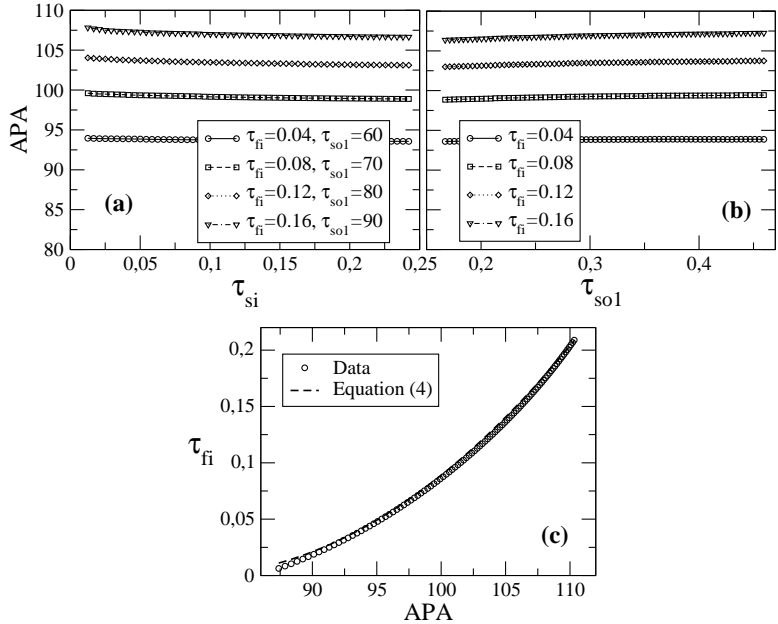}
\caption{\protect
  (a) Amplitude $\hbox{APA}$ as a function of $\tau_\textrm{si}$ for four
    different pairs of fixed values $\tau_\textrm{fi}$ and
    $\tau_\textrm{so1}$.
  (b) Dependence of the amplitude $\hbox{APA}$ on 
    time $\tau_\textrm{so1}$ for $\tau_\textrm{si}=10.7$~ms
    and four different values of $\tau_\textrm{fi}$.
    (c) Time $\tau_\textrm{fi}$ as a function of APA
    for $\tau_{\rm si}=10.7$~ms and $\tau_{\rm so1}=73.7$~ms.}
\label{fig03}
\end{figure}
%%%%%%%%%%%%%%%%%%%%%%%%%%%%%%%%%%%%%%%%%%%%%%%%%%%%%%%%%%%%%%%%%%%%%%

The paper is organized as follows. In Section \ref{sec:model} we
shortly summarize the BOCF model and discuss the role of
the three fit parameters that we selected to model specific AP.
In Section \ref{sec:map} we show how these parameters
can be adjusted to obtain a a faithful representation of the AP
properties $\hbox{APA}$, APD$_n$, and
in Section \ref{sec:optimal} we demonstrate the specific AP modeling
for surrogate data generated with the 
CRN model \cite{Courtemanche/etal:1998}.
A summary of our main
findings and discussion of their relevance is given
in Section \ref{sec:conclusions} 
In the Appendix, we provide analytical calculations
for the BOCF model that motivated our choice of fit parameters for the
AP modeling.

%%%%%%%%%%%%%%%%%%%
\section{BOCF model for atrial physiology}
\label{sec:model}

The BOCF model has four state variables, which are the scaled TMV $u$,
and three variables $v$, $w$ and $s$ describing the gating of
(effective) net currents through the cell membrane. 
The TMV $V$ is obtained from $u$ via the linear relation
$V=V_R(1+\alpha u)$, where for atrial tissue we set $V_R=-84.1$~mV
for the resting potential and $\alpha=1.02$~\cite{Lenk/etal:2015}.
The time-evolution of $u$ is given by the reaction-diffusion equation
\begin{equation}
\partial_{t}u = D\Delta u + J(u,v,w,s)+J_\textrm{stim}\,, \label{eq:BOdu}
\end{equation}
where $J=J(u,v,w,s)$ is the total ionic current and $J_\textrm{stim}$ an
external stimulus current. 
For modeling  of single-cell action potentials, as considered in this work, we set $D=0$. 
The total ionic current decomposes into three net currents,
a fast inward sodium current $J_{\rm fi}=J_{\rm fi}(u,v)$, a slow inward
calcium current $J_{\rm si}(u,w,s)$, and a slow  outward 
potassium current $J_{\rm so}=J_{\rm so}(u)$,
\begin{equation}
J(u,v,w,s)= J_{\rm fi}(u,v) + J_{\rm si}(u,w,s)+ J_{\rm so}(u)\,.
\label{eq:J}
\end{equation}
These currents are controlled by the gating variables, which evolve 
according to
\begin{equation}
\partial_{t}(v,w,s) = (E(u,v),F(u,w),G(u,s))\,, \label{eq:BOdvws}%
\end{equation}
where the nonlinear functions $F$, $G$ and $H$,
are specified in Section \ref{sec:appendmodel}.
There we show that the four differential equations
(\ref{eq:BOdu}) and (\ref{eq:BOdvws}) can be reduced to a system of two
differential equations.
This reduction 
shows that the three characteristic times $\tau_{\rm fi}$, $\tau_{\rm si}$ and
$\tau_{\rm so1}$, which fix the typical duration of the respective currents,
govern the shape of the AP [cf.~Eq.~(\ref{uR1}) in the Appendix].
We take these three time scales
as parameters for fitting a specific AP and keep all other parameters
fixed. For the values of the fixed parameters we here consider the
set determined for
the electrically remodeled tissue due to atrial
fibrillation \cite{Seemann/etal:2010, Lenk/etal:2015}.

%%%%%
\section{Parameter dependence of BOCF action potentials}
\label{sec:map}

In this section we show that in the BOCF model
the amplitude $\hbox{APA}$ can be expressed by a quadratic polynomial
of the times $\tau_{\rm fi}$, and the APD$_n$ by
cubic polynomials of $\tau_{\rm si}$ and $\tau_{\rm so1}$.

The dependence of $\hbox{APA}$ and the APD$_n$ on the
characteristic times, was determined from
generated AP in single-cell simulations of
the BOCF model by applying periodically,
with a frequency $f=3$ Hz, a constant stimulus current
of $40$~pA, corresponding to an amplitude of 
$4.76$~s$^{-1}$ for the current $J_{\rm stim}$ in Eq.~(\ref{eq:BOdu}),
%$\tilde{J}_{\rm stim}=40$~pA
for a time period of $3.5$~ms.
%During this time period its %value The
%amplitude was
%$j_\textrm{stim}=4.76~s^{-1}$, % in Eq.~(\ref{eq:BOdu}),
%corresponding to 40~pA in SI units.
The resulting time evolution of the TMV in response to this stimulus
was calculated by integrating Eqs.~(\ref{eq:BOdu}) and
(\ref{eq:BOdvws}) for the initial conditions $u_0=0$, $v_0=1$,
$w_0=1$ and $s_0=0$.
This was done for
$(\tau_\textrm{fi}, \tau_\textrm{si}, \tau_\textrm{so1})\in [0.002,0.210]\times [5.9,22.4]\times [40,110]$ (in ms) with a resolution
$\Delta \tau_{\rm fi} = 0.0021$~ms ($100$ values),
$\Delta \tau_{\rm si} = 0.3$~ms ($56$ values) and
$\Delta \tau_{\rm so1} = 1$~ms ($71$ values).
The AP was recorded after a transient time of 10~s. 
%In the following, all times will be given in milliseconds unless noted
%otherwise.

As shown for a few representative pairs of fixed
values of  $\tau_{\rm fi}$ and $\tau_{\rm so1}$ in
Fig.~\ref{fig03}(a) and \ref{fig03}(b),
the APA depends only very weakly on $\tau_{\rm si}$ and $\tau_{\rm so1}$.
Neglecting these weak dependencies, on $\tau_{\rm si}$ and $\tau_{\rm so1}$,
we find the $\hbox{APA}$ to increase monotonically with $\tau_{\rm fi}$
in the range $[85,110]$~mV relevant for
human atria.   % \textbf{Zitat?}.}
In Fig.~\ref{fig03}(c) we show that
the parameter $\tau_{\rm fi}$
can be well described by the quadratic polynomial 
%$\mathcal{V}$
\begin{equation}
  \tau_{\rm fi} = c_0 \hbox{APA}^2 + c_1 \hbox{APA} + c_2 \, ,
\label{taufi-amplitude}
\end{equation}
where the coefficients $c_i$ and the coefficient of determination $R^2$ of
the fit are given in Table \ref{tab02}.
%with $c_0=2.35\pm 0.06$ ($\times 10^{-4}\hbox{\ ms}/\hbox{(mV)}^2$),
%$c_1=-3.8\pm 0.1$ ($\times 10^{-2}\hbox{\ ms}/\hbox{mV}$) and
%$c_2=1.52\pm 0.05$ (ms). The fit yields $R^2=0.9996$.

\begin{figure}[htb]
\centering
\includegraphics[width=0.9\textwidth]{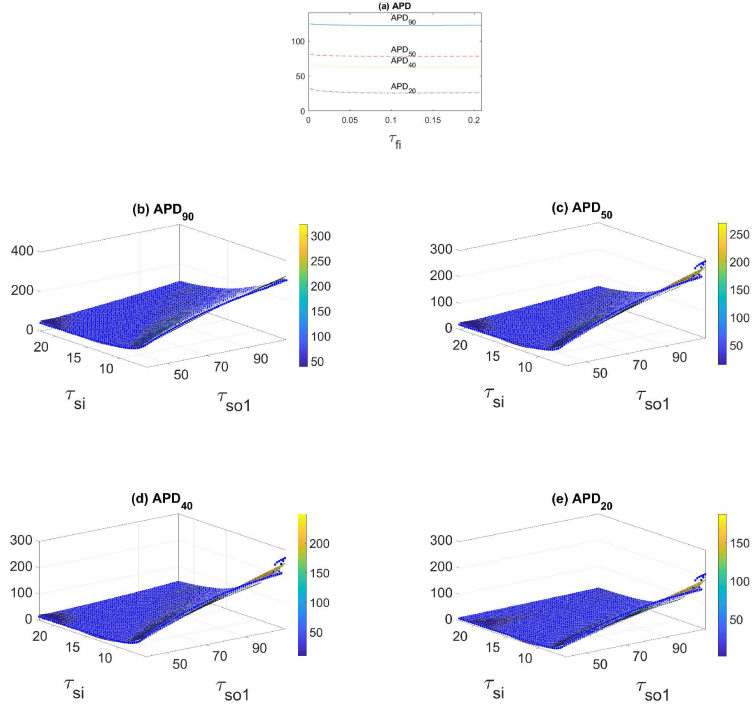}
\caption{\protect
  (a) APD$_n$ as a function of $\tau_\textrm{fi}$ for a pair of fixed
  values $\tau_\textrm{si}=10.7$~ms and
  $\tau_\textrm{so1}=73.675$~ms.  
  (b)-(e) Dependence of the APD$_n$ on
  $\tau_{\rm si}$ and $\tau_{\rm so1}$ for fixed $\tau_\textrm{fi}=0.0835$~ms.
  The meshes of points (black bullets)
  indicate the simulation results, and the surfaces
  refer to the fits of the meshes, according to Eq.~(\ref{surface}).
  All quantities are given in ms.}
%\textbf{Kann man numerische Daten und
%  Fitdaten vom Polynom unterscheiden?}}
\label{fig04}
\end{figure}

Likewise, as demonstrated in Fig.~\ref{fig04}(a) for one fixed pair
of values  of  $\tau_{\rm si}$ and $\tau_{\rm so1}$, the APD$_n$ 
are almost independent of $\tau_\textrm{fi}$. 
Their dependence on $\tau_{\rm si}$ and $\tau_{\rm so1}$,
shown in Figs.~\ref{fig04}(b)-(e),
can be well fitted by the polynomials
\begin{equation}
  \hbox{APD}_n(\tau_{\rm si},\tau_{\rm so1}) =
  \sum_{m=0}^3 \sum_{k=0}^{3-m} c^{(n)}_{mk}\tau_{\rm si}^m \tau_{\rm so1}^k \, .
\label{surface}
\end{equation}
where the coefficients $c^{(n)}_{mk}$ are 
listed in Table \ref{tab02} together with the $R^2$ values of the fits.

%%%%%%%%%%%%%%%%%%%%%%%%%%%%%%%%%%%%%%%%%%%%%%%%%
\begin{table}[!t]
\caption{\protect
  Polynomial coefficients and $R^2$ values of the fits of
  APA to Eq.~(\ref{taufi-amplitude}) and of
  the surfaces
  APD$_n(\tau_{\rm si}, \tau_{\rm so1})$ 
  to %Eqs.~(\ref{surface1}) and
  Eq.~(\ref{surface}). The values of coefficients $c_{mk}^{(n)}$ are given in
  units of $\hbox{mV}/\hbox{(ms)}^{m+k}$.}
\begin{center}
\begin{tabular}{|c|c||c|c|c|c|c|}
\hline			
    {\bf Coeffs.} & $\mathbf{APA}$ & {\bf Coeffs.} &
    $\mathbf{APD_{90}}$ & $\mathbf{APD_{50}}$ & $\mathbf{APD_{40}}$ & $\mathbf{APD_{20}}$ \\
    Eq.~(\ref{taufi-amplitude}) &  & Eq.~(\ref{surface}) & & & & \\    
\hline
$c_0$ & 2.35 &
$c^{(n)}_{00}$ 		& $98$ 	& $85$ 	& $84$ 	& $ 82$\\
$\pm\Delta c_0$ & $\pm 0.06$ &
$\pm \Delta c^{(n)}_{00}$ & $\pm 10$ 	& $\pm 10$ 	& $\pm 10$ 	& $\pm  10$\\
\hline
$c_1$ & -3.8 &
$c^{(n)}_{10}$ 		& $5.4$ 		& $5.0$ 		& $ 4.7$ 		& $3.8$\\
$\pm\Delta c_1$ & $\pm 0.1$ &
$\pm \Delta c^{(n)}_{10}$ & $\pm 0.3$ & $\pm 0.3$ & $\pm 0.4$ 	& $\pm 0.3$\\
\hline
$c_2$ & 1.52 &
$c^{(n)}_{01}$ 		& $-33$ 	& $-33$ 	& $-33$ 	& $-32$\\
$\pm \Delta c_2$ & $\pm 0.05$ &
$\pm \Delta c^{(n)}_{01}$ & $\pm 1$ 	& $\pm 1$ 	& $\pm 1$ 	& $\pm 1$\\
\hline
$\mathbf{R^2}$ & $\mathbf{0.9996}$ & 
$c^{(n)}_{20}$ 		& $0.0001$ 	& $-0.0010$ 	& $0.0001$ 	& $0.003$\\
\cline{1-2}
\multicolumn{2}{c|}{\ } &
$\pm \Delta c^{(n)}_{20}$ & $\pm 0.004$ & $\pm 0.004$ & $\pm 0.004$ & $\pm 0.004$\\
\cline{3-7}
%\hline
\multicolumn{2}{c|}{\ } &
$c^{(n)}_{11}$ 		& $-0.40$ 	& $-0.41$ 	& $-0.41$ 	& $-0.43$\\
\multicolumn{2}{c|}{\ } &
$\pm \Delta c^{(n)}_{11}$ & $\pm 0.01$ & $\pm 0.01$ & $\pm 0.01$ & $\pm 0.01$\\
\cline{3-7}
%\hline
\multicolumn{2}{c|}{\ } &
$c^{(n)}_{02}$ 		& $2.47$ 	& $2.56$ 	& $2.61$ 	& $2.85$\\
\multicolumn{2}{c|}{\ } &
$\pm \Delta c^{(n)}_{02}$ & $\pm 0.06$ & $\pm 0.06$ & $\pm 0.07$ & $\pm 0.06$\\
\cline{3-7}
%\hline
\multicolumn{2}{c|}{\ } &
$c^{(n)}_{30}$ 		& $-0.0000721$ 	& $-0.00005$ 	& $-0.00004$ 	& $-0.00002$\\
\multicolumn{2}{c|}{\ } &
$\pm \Delta c^{(n)}_{30}$ & $\pm 0.00002$ & $\pm 0.00002$ & $\pm 0.00002$ & $\pm 0.00002$\\
\cline{3-7}
%\hline
\multicolumn{2}{c|}{\ } &
$c^{(n)}_{21}$ 		& $0.0012591$ 	& $0.00096$ 	& $0.00079$ 	& $0.00018$\\			
\multicolumn{2}{c|}{\ } &
$\pm \Delta c^{(n)}_{21}$ & $\pm 0.00007$ & $\pm 0.00007$ & $\pm 0.00007$ & $\pm 0.00007$\\
\cline{3-7}
%\hline
\multicolumn{2}{c|}{\ } &
$c^{(n)}_{12}$ 		& $0.0027$ 	& $0.0045$ 	& $ 0.0057$ 	& $0.0103$\\
\multicolumn{2}{c|}{\ } &
$\pm\Delta c^{(n)}_{12}$ & $\pm 0.0003$ & $\pm 0.0003  $ & $\pm 0.0003$ & $\pm 0.0003$\\
\cline{3-7}
%\hline
\multicolumn{2}{c|}{\ } &
$c^{(n)}_{03}$ 		& $-0.045$ 	& $-0.050$ 	& $-0.053$ 	& $-0.069$\\
\multicolumn{2}{c|}{\ } &
$\pm\Delta c^{(n)}_{03}$ & $\pm 0.001$ & $\pm  0.001$ & $\pm 0.001$ & $\pm 0.001$\\
\cline{3-7}
%\hline	
\multicolumn{2}{c|}{\ } &
$\mathbf{R^2}$ & $\mathbf{0.9956}$ & $\mathbf{0.9938}$ & $\mathbf{0.9926}$ & $\mathbf{0.9866}$\\
\cline{3-7}
%\hline		
\end{tabular}
\end{center}
\label{tab02}
\end{table}
%%%%%%%%%%%%%%%%%%%%%%%%%%%%%%%%%%%%%%%%%%%%%%%%
%%%%%%%%%%%%%%%%%%%%%%%%%%%%%%%%%%%%%%%%%%%%%%%%%%%%%%%%%%%%%%%%%%%%%%
%%%%%%%%%%%%%%%%%%%%%%%%%%%%%%%%%%%%%%%%%%%%%%%%%%%%%%%%%%%%%%%%%%%%%%
%%%%%%%%%%%%%%%%%%%%%%%%%%%%%%%%%%%%%%%%%%%%%%%%%%%%%%%%%%%%%%%
\begin{figure}[!t]
\centering
\includegraphics[width=0.95\textwidth]{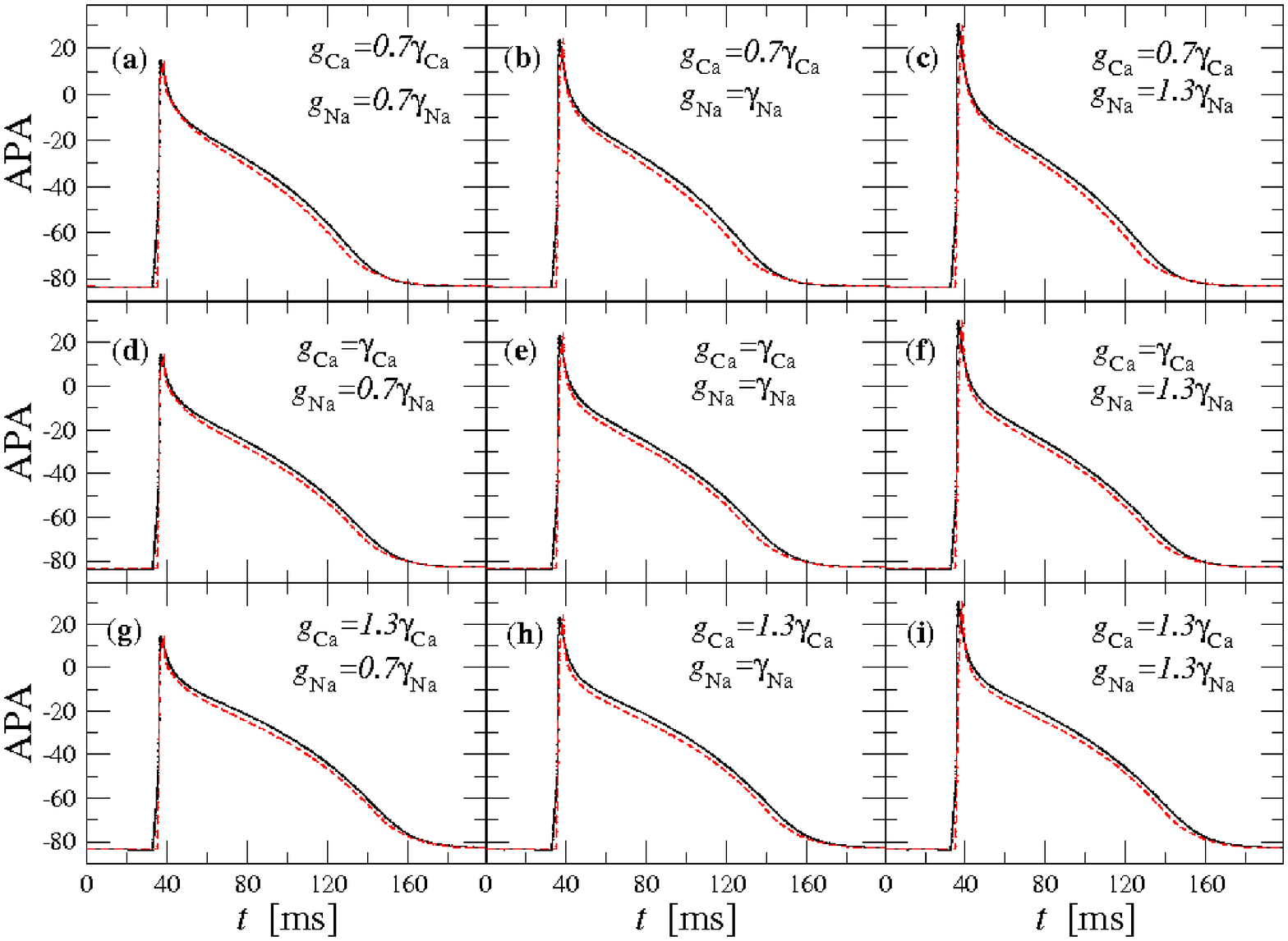}
\caption{\protect 
  Nine surrogate AP generated with the CRN model (solid lines) for
  different $g_{\rm Na}$ and $g_{\rm Ca}$
  in comparison with the corresponding AP modeled with the BOCF model
  (dashed lines).
  The reference values are the ones corresponding to the remodeling
  case, namely $\gamma_{\rm Ca}=0.0433$ nS$/$ps and $\gamma_{\rm Na}=7.8$
  nS$/$ps.}
\label{fig05}
\end{figure}
%%%%%%%%%%%%%%%%%%%%%%%%%%%%%%%%%%%%%%%%%%%%%%%%%%%%%%%%%%%%%%%

%%%%%%%%%%%%
\section{Modeling of patient-specific action potentials with the BOCF model}
\label{sec:optimal}

Let us denote by $\mathcal{V}$ the APA and by
$\mathcal{D}_n$ the values of the APD$_n$ of a
specific patient.
To model the corresponding AP with the BOCF model, we determine
$\tau_{\rm fi}$ by inserting $\hbox{APA}=\mathcal{V}$
in Eq.~(\ref{taufi-amplitude}) and 
$(\tau_\textrm{si}, \tau_\textrm{so1})$ by minimizing
the sum of the squared deviations between the the APD$_n$, i.~e.\ the
function
\begin{equation}
\mathcal{F}(\tau_{\rm si},\tau_{\rm so1}) =
\sum_{n}
\big [ \hbox{APD}_n(\tau_{\rm si},\tau_{\rm so1})-\mathcal{D}_n \big ]^2\,.
\label{functional}
\end{equation}
For the numerical procedure we used the Levenberg-Marquardt
algorithm~\cite{recipes}.
As one sees from Figs.~\ref{fig04}(b)-(e),
the APD vary monotonically with the
time scales in the ranges fixed above.
We checked that the Hessian is positive definite in the corresponding
region, implying unique solutions when minimizing $\mathcal{F}$.

To demonstrate the adaptation procedure, we generated surrogate
AP with the CRN model\cite{Courtemanche/etal:1998} for electrically
remodeled tissue due to atrial fibrillation \cite{Seemann/etal:2010}.
Specifically, we consider the maximal conductances,
$g_{\rm Ca}$ and $g_{\rm Na}$ of the calcium and
sodium currents to vary, while keeping all other
parameters fixed to the values corresponding to the
electrically remodeled tissue.
%\textbf{Fest gew\"ahlt entsprechend des Seemann-Parametersatzes
%f\"ur das CRN-Modell bei elektrisch remodelliertem Gewebe aufgrund
%Vorhofflimmerns?}
The conductance $g_{\rm Ca}$ 
%calcium current is a slow inward current of the $L$-type that
affects both the AP plateau and the repolarization phase and the
$g_{\rm Na}$ controls mainly the amplitude of the 
AP~\cite{Courtemanche/etal:1998}.
%\textbf{Der letzte Satz bestand urspr\"unglich aus zwei S\"atzen, wovon ersterer syntaktisch kein Satz war. Ich habe versucht zu verstehen, was gemeint war. Bitte ggf.\ korrigieren.}

Figure~\ref{fig05} shows nine examples of AP generated with the CRN model,
which cover a wide range of APA and APD.
In Figs.~\ref{fig05}(a)-(e) we allow $g_{\rm Na}$ and $g_{\rm Ca}$ to differ
by factors between $70\%$ and $130\%$ from their values
$\gamma_{\rm Na}=7.8$~nS/pF 
and $\gamma_{\rm Ca}=0.0433$~nS/pF for the electrically remodelled
tissue\cite{Seemann/etal:2010}.
The corresponding AP modeled with the BOCF, i.~e.\ for
$\tau_{\rm fi}$ from Eq.~(\ref{taufi-amplitude}), and
$\tau_{\rm si}$ and $\tau_{\rm so1}$ obtained from the minimization of 
$\mathcal{F}(\tau_{\rm si},\tau_{\rm so1})$
in Eq.~(\ref{functional}), are shown as dashed lines in the figures.
In all cases these reproduce well the AP shapes generated with the CRN model.

%%%%%%%%%%%%%%%%%%%%%%%%%%%%%%%%%%%%%%%%%%%%%%%%%%%%%%%%%%%%%%%
\begin{table}[!t]
  \caption{\protect
    APA $\mathcal{V}$ and APD$_n$ values $\mathcal{D}_n$ for
    the examples shown in Fig.~\ref{fig05}. The $\Delta$ values
    give the deviations of the individual form parameters according
    to Eq.~(\ref{delta}) and $\Delta\mathcal{A}$ is the deviation
    between both AP based in their $L_2$-norms, as defined in
    Eq.~(\ref{l2norm}).}
\begin{center}
\begin{tabular}{|c|c|c|c|c|c|c|c|}
\cline{3-8}
\multicolumn{2}{c|}{}
  & $\mathcal{D}_{90}$ & $\mathcal{D}_{50}$ & $\mathcal{D}_{40}$ & $\mathcal{D}_{20}$ & $\mathcal{V}$ & $\Delta\mathcal{A}$ \\
\multicolumn{2}{c|}{}
  & (ms) & (ms) & (ms) & (ms) & (mV) & ($\times 10^{-2}$) \\\hline 
\multirow{2}{0.3cm}{\rotatebox{90}{Fig.\ref{fig05}a\ }}
      & {\scriptsize CRN}
      & 107.7
      & 66.36
      & 53.07
      & 21.77
      & 98.43
      &   \\\cline{2-7}
      & {\scriptsize BOCF}
      & 102.9
      & 61.44
      & 48.09
      & 19.26
      & 98.15
      & {\bf 4.5}  \\\cline{2-7}
      & $\Delta$ ($\times 10^{-2}$)
      & {\bf 4.5}
      & {\bf 7.4}
      & {\bf 9.4}
      & {\bf 11.5}
      & {\bf 0.3}
      &   \\\hline
\multirow{2}{0.3cm}{\rotatebox{90}{Fig.\ref{fig05}b\ }}
      & {\scriptsize CRN}
      & 106.8
      & 66.02
      & 53.19
      & 23.21
      & 107.14
      &   \\\cline{2-7}
      & {\scriptsize BOCF}
      & 101.1
      & 60.75
      & 47.73
      & 19.10
      & 107.0
      & {\bf 6.0}  \\\cline{2-7}
      & $\Delta$ ($\times 10^{-2}$)
      & {\bf 5.3}
      & {\bf 8.0}
      & {\bf 10.3}
      & {\bf 17.7}
      & {\bf 0.15}
      &   \\\hline
\multirow{2}{0.3cm}{\rotatebox{90}{Fig.\ref{fig05}c\ }}
      & {\scriptsize CRN}
      & 105.9
      & 65.59
      & 53.08
      & 24.04
      & 114.1
      &   \\\cline{2-7}
      & {\scriptsize BOCF}
      & 100.25
      & 60.71
      & 48.02
      & 19.99
      & 112.9
      & {\bf 7.3} \\\cline{2-7}
      & $\Delta$ ($\times 10^{-2}$)
      & {\bf 5.3}
      & {\bf 7.4}
      & {\bf 9.5}
      & {\bf 16.8}
      & {\bf 1.0}
      &   \\\hline
\multirow{2}{0.3cm}{\rotatebox{90}{Fig.\ref{fig05}d\ }}
      & {\scriptsize  CRN}
      & 115.7
      & 72.57
      & 59.03
      & 26.03
      & 98.44
      &   \\\cline{2-7}
      & {\scriptsize BOCF}
      & 110.5
      & 68.02
      & 53.85
      & 21.59
      & 98.18
      & {\bf 4.6} \\\cline{2-7}
      & $\Delta$ ($\times 10^{-2}$)
      & {\bf 4.5}
      & {\bf 6.3}
      & {\bf 8.8}
      & {\bf 17.1}
      & {\bf 0.26}
      &   \\\hline
\multirow{2}{0.3cm}{\rotatebox{90}{Fig.\ref{fig05}e\ }}
      & {\scriptsize CRN}
      & 114.3
      & 71.75
      & 58.67
      & 27.18
      & 107.1
      &   \\\cline{2-7}
      & {\scriptsize BOCF}
      & 108.3
      & 66.94
      & 53.17
      & 21.40
      & 107.0
      & {\bf 5.9} \\\cline{2-7}
      & $\Delta$ ($\times 10^{-2}$)
      & {\bf 5.3}
      & {\bf 6.7}
      & {\bf 9.4}
      & {\bf 21.3}
      & {\bf 0.08}
      &   \\\hline
\multirow{2}{0.3cm}{\rotatebox{90}{Fig.\ref{fig05}f\ }}
      & {\scriptsize CRN}
      & 113.2
      & 71.08
      & 58.30
      & 27.84
      & 113.9
      &   \\\cline{2-7}
      & {\scriptsize BOCF}
      & 107.3
      & 66.64
      & 53.21
      & 22.24
      & 112.9
      & {\bf 7.1} \\\cline{2-7}
      & $\Delta$ ($\times 10^{-2}$)
      & {\bf 5.3}
      & {\bf 6.2}
      & {\bf 8.7}
      & {\bf 20.1}
      & {\bf 0.9}
      &   \\\hline
\multirow{2}{0.3cm}{\rotatebox{90}{Fig.\ref{fig05}g\ }}
      & {\scriptsize CRN}
      & 124.5
      & 81.00
      & 67.44
      & 31.92
      & 98.24
      &   \\\cline{2-7}
      & {\scriptsize BOCF}
      & 119.6
      & 76.76
      & 62.02
      & 25.66
      & 98.00
      & {\bf 4.9} \\\cline{2-7}
      & $\Delta$ ($\times 10^{-2}$)
      & {\bf 3.9}
      & {\bf 5.2}
      & {\bf 8.0}
      & {\bf 19.6}
      & {\bf 0.2}
      &   \\\hline
\multirow{2}{0.3cm}{\rotatebox{90}{Fig.\ref{fig05}h\ }}
      & {\scriptsize CRN}
      & 122.6
      & 79.69
      & 66.56
      & 32.82
      & 106.9
      &   \\\cline{2-7}
      & {\scriptsize BOCF}
      & 117.0
      & 75.25
      & 60.92
      & 25.42
      & 106.9
      & {\bf 6.2} \\\cline{2-7}
      & $\Delta$ ($\times 10^{-2}$)
      & {\bf 4.6}
      & {\bf 5.6}
      & {\bf 8.5}
      & {\bf 22.5}
      & {\bf 0.02}
      &   \\\hline
\multirow{2}{0.3cm}{\rotatebox{90}{Fig.\ref{fig05}i\ }}
      & {\scriptsize CRN}
      & 121.2
      & 78.68
      & 65.86
      & 33.30
      & 113.7
      &   \\\cline{2-7}
      & {\scriptsize BOCF}
      & 115.6
      & 74.59
      & 60.63
      & 26.63
      & 112.8
      & {\bf 7.5} \\\cline{2-7}
      & $\Delta$ ($\times 10^{-2}$)
      & {\bf 4.7}
      & {\bf 5.2}
      & {\bf 7.9}
      & {\bf 21.2}
      & {\bf 0.8}
      &   \\\hline
\end{tabular}
\end{center}
\label{tab01}
\end{table}
%%%%%%%%%%%%%%%%%%%%%%%%%%%%%%%%%%%%%%%%%%%%%%%%%%%%%%%%%%%%%%%%%%%%%%%

To quantify the difference between the AP,
we denote by
$\mathcal{A}_{\rm CRN}(t)$ and  $\mathcal{A}_{\rm BOCF}(t)$
their time course, and compute their relative deviation based on the
$L_2$-norm,
\begin{equation}
  \Delta\mathcal{A} = \frac{\vert\vert \mathcal{A}_{\rm BOCF}(t) - \mathcal{A}_{\rm CRN}(t) \vert\vert_{L_2}}{\vert\vert \mathcal{A}_{\rm CRN}(t) \vert\vert_{L_2}} \, ,
\label{l2norm}
\end{equation}
where
\begin{equation}
  \vert\vert \mathcal{A}(t) \vert\vert_{L_2} \equiv
  \left ( \int_{t_i}^{t_f} \mathcal{A}^2(t)dt \right )^{1/2}  \, .
\end{equation}
The initial time $t_i$ and final time $t_f$ are defined as the
times for which $u(t_i)=u(t_f)=\theta_0$ with
$\theta_0=0.015473$ (see Appendix), with opposite signs of the
corresponding time derivatives, i.e.~$\tfrac{du}{dt}\vert_{t_i}>0$
and $\tfrac{du}{dt}\vert_{t_f}<0$.

For the examples in Fig.~\ref{fig05}, Table \ref{tab01} gives
the values of APA and the APD$_n$ for surrogate AP generated
with CRN model and the adapted BOCF model, together with the
deviations $\Delta\mathcal{A}$.
The largest differences between both AP correspond to deviations
of the order of $5\%$ to $7\%$.

The relative errors of the APA and APD$_n$ 
\begin{equation}
  \Delta = \frac{\vert X_{\rm BOCF}-X_{\rm CRN} \vert}{X_{\rm CRN}} \, ,
\label{delta}
\end{equation}
with $X$ representing either $\mathcal{V}$ or $\mathcal{D}_n$ are
also given in Table \ref{tab01}.
The APA show deviations up to $1\%$ and the APD$_n$ up to around $10\%$
for all $n$ except $20$. The APD$_{20}$ refers to the TMV level
closest to the maximum and exhibits larger deviations of about $20\%$
for even small shape deviations.
%As for the APDs, all of them have relative errors of the order of
%less than $10\%$, except the APD$_{20}$, which, since is the closest
%to the AP maximum, is the most sensitive one to shape deviations. 

%In order to ascertain if this matching can be considered as comparably
%good to fittings of APs with established models reported in the
%literature\cite{Wilhelms/etal:2013}.
%We take the known typical amplitude and APDs reproduced with standard
%models (see Tab.~1 and 2 in Ref.~\cite{Wilhelms/etal:2013})
%and compare their average value $\mu_{\rm models}$ and standard deviation
%$\sigma_{\rm models}$ with the corresponding deviations between both models.
%To that end, we introduce the quantity
%\begin{equation}
%\Delta = \frac{\Delta_B}{\Delta_0},
%\end{equation}
%with
%\begin{subequations}
%\begin{eqnarray}
%\Delta_B &=& \frac{\vert X_{\rm CRN}-X_{\rm BOCF}
%\vert}{X_{\rm CRN}} \, , \label{deltaB}\\
%\Delta_0 &=& \frac{\sigma_{\rm model}}{\mu_{\rm model}} \, , \label{delta0} 
%\end{eqnarray}
%\label{delta}%
%\end{subequations}
%where $X$ represents either the amplitude or one of the APD$_n$.
%Values of $\Delta$ below unit mean that the deviations observed
%in our examples are typically smaller than the deviations observed
%in accepted models of AP.
%\textbf{Ist eigentlich der Vergleich zwischen den verschiedenen Modellen
%in der Arbeit von Wilhelms et al. bezogen auf ein spezifisches AP
%anzusehen, d.~h.\ kann man diesen Vergleich wirklich analog
%betrachten zu dem, was wir hier einordnen m\"ochten?}

%%%%%%%% 
\section{Conclusions}
\label{sec:conclusions}

In this work we showed how to model patient-specific action potentials
by adjusting three characteristic time scales, which are associated with
the net sodium, calcium and potassium ionic currents.
The framework explores the possibilities of 
parameter adjustment of an atrial physiology model, namely the
BOCF model\cite{Bueno-Orovio/etal:2008}, to reproduce
AP shapes with a given amplitude, width and duration.
The BOCF model is defined through a reaction-diffusion equation,
coupled to three equations for gating variables
that describe the opening and closing of ionic channels.
It is simple enough to guarantee
low computational costs for even extensive simulations of spatio-temporal dynamics
\cite{Richter:2017}.
Through a semi-analytical approach given in the Appendix
we showed why the three
ionic currents 
suffice to derive the main features of empirical AP.

The high flexibility for case-specific applications can
be used for clinical purposes.
Using the optimization procedure for AP shape adjustment, 
the three characteristic times are retrieved, 
which are directly connected to the ion-type specific net currents.
AP shapes showing pathological features will be reflected in the values of one (or more) times outside acceptable
ranges. Accordingly, one can associate a corresponding net current
and therefore identify the class of membrane currents, where
pathologies should be present. 
In this sense the clinical diagnosis can be supported by the modeling.

Furthermore, in case information is obtained about AP shapes from different
places of the atria, e.~g.\ by using a lasso catheter, a corresponding AP
shape modeling would allow one to construct a patient-specific model with
spatial heterogeneities.
Based on this, it could become possible to generate spatio-temporal activation
pattern and to identify possible pathologies associated in the dynamics of
the action potential propagation.

%For the optimized adjustment,
%we defined the shape of the AP with four AP durations at
%different potential levels and map them, together with the AP
%amplitude, into three single time scales, each one corresponding to
%one ionic channel in the simplified (BOCF) model.
%The analysis shown in this paper covers a wide range of APs,
%with different amplitudes, durations and shapes.
%
%As an overall result, we have shown that APs can be well reproduced
%by this optimal adjustment.
%The high flexibility for case-specific applications is now established
%for using it in the clinical context.
%A natural application that follows
%from the results in this paper is to adjust the BOCF model to concrete
%physical properties of real patients.
%Empirical APs are intra-atrial signals that can be recorded for
%specific patients.
%In the clinical context, our framework provides valuable insight,
%since it uncovers the time scales of main ionic currents, and detecting
%which of them lie outside the normal values is an important step for
%reliable diagnostics.
%Further tests against empirical data are however needed, an open issue
%that will be presented elsewhere.

%\section{Supporting information}

\appendix
%%%%%%%%%%
\section{Appendix:
            Dynamical features of the BOCF model: a semi-analytical approach.}
\label{sec:appendmodel}

%%%%%%%%%%%%%%%%%%%%%%%%%%%%%%%%%%%%%%%%%%%%%%%%%%%%%%%%%%%%%%%%%%%%%%%%
\begin{figure}[!b]
\centering
\includegraphics[width=0.6\textwidth]{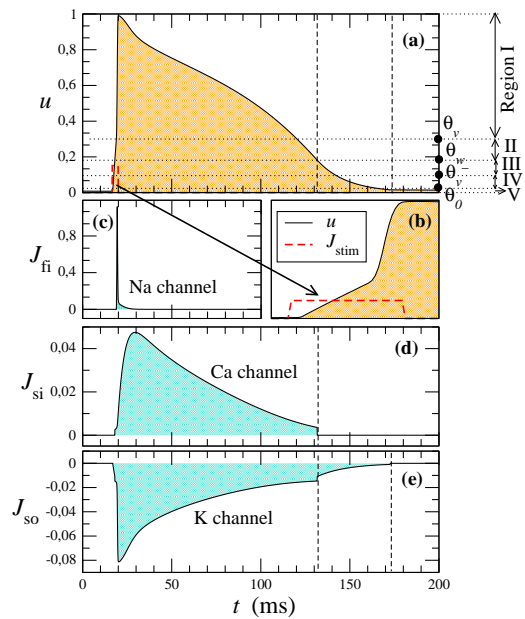}
\caption{\protect
  Time evolution of one AP together with each ionic current:
  (a) AP variable $u$ with the stimulus current $J_{\rm stim}$,
  with (b) a close-up for a time interval of $3.5$ ms.
  Vertical dashed lines intersect the AP at one specific dotted line,
  thus bounding the time intervals corresponding to each region of
  $u$-values (see text).
  The ionic currents correspond to 
  (c) the Na channel ($J_{\rm fi}$),
  (d) the Ca channel ($J_{\rm si}$), and
  (e) the K channel ($J_{\rm so}$),
  see Eqs.~(\ref{eq:BOdvws}) and (\ref{allJ}).
  All currents are given in (ms)$^{-1}$.}
\label{fig06}
\end{figure}
%%%%%%%%%%%%%%%%%%%%%%%%%%%%%%%%%%%%%%%%%%%%%%%%%%%%%%%%%%%%%%%%%%%%%%%%
%%%%%%%%%%%%%%%%%%%%%%%%%%%%%%%%%%%%%%%%%%%%%%%%%%%%%%%%%%%%%%%%%%%%%%%%
\begin{figure}[!t]
\centering
\includegraphics[width=0.9\textwidth]{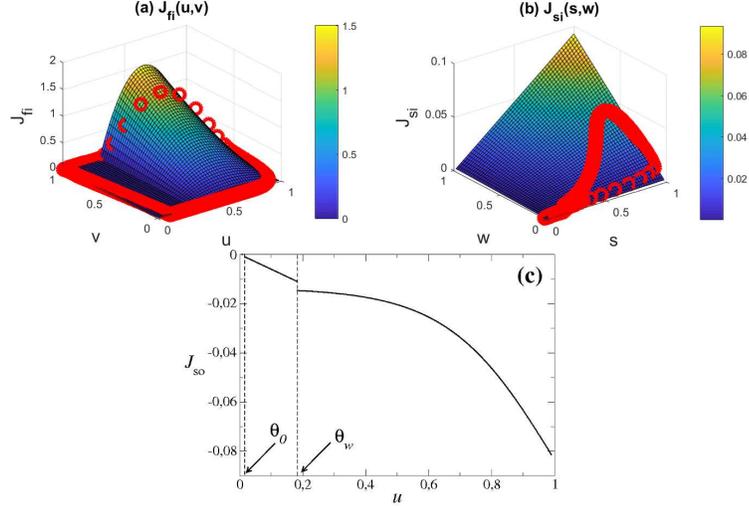}
\caption{\protect 
  Ionic current (a) $J_{\rm fi}$ and (b) $J_{\rm si}(u,w)$ as a
  function of the respective gating variables.
  The red circles indicate the path corresponding to 
  Eqs.~(\ref{eq:BOdu}) and (\ref{eq:BOdvws}) and sketched
  in Fig.~\ref{fig06} as a function of time.
  (c) Dependence of ionic current $J_{\rm so}$ on variable $u$.}
\label{fig07}
\end{figure}
%%%%%%%%%%%%%%%%%%%%%%%%%%%%%%%%%%%%%%%%%%%%%%%%%%%%%%%%%%%%%%%%%%%%%%%%
%%%%%%%%%%%%%%%%%%%%%%%%%%%%%%%%%%%%%%%%%%%%%%%%%%%%%%%%%%%%%%%%%%%%%%%%
\begin{figure}[!t]
\centering
\includegraphics[width=0.7\textwidth]{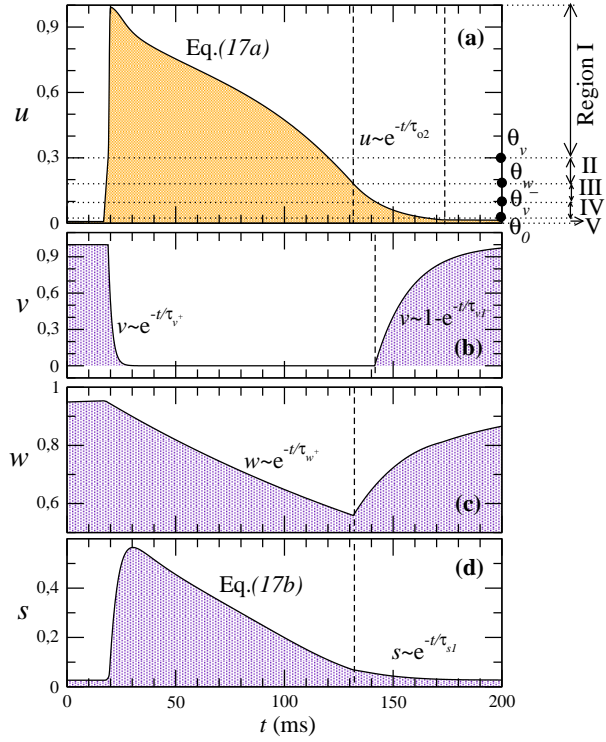}
\caption{\protect 
  Time evolution of the four variables of the BOCF model:
  {\bf (a)} AP variable $u$ and the three gating variables
  {\bf (b)} $v$,
  {\bf (c)} $w$ and
  {\bf (d)} $s$.
  The horizontal dotted lines in (a) indicate the ranges of $u$-values,
  where the evolution of the set of variables changes discontinuously.
  Vertical dashed lines intersect the AP at one specific dotted line,
  thus bounding the time intervals corresponding to each region of
  $u$-values.
  In several of such time intervals, some of the variables
  decay exponentially and independently from the other variables,
  which simplifies the model considerably.
  In the regions where no exponential evolution is indicated the model
  follows the reduced system of equations in (\ref{R1}):
  in plot (a) one sees the integration of Eq.~(\ref{uR1}) and in plot (d)
  the integration of Eq.~(\ref{sR1}).}
\label{fig08}
\end{figure}
%%%%%%%%%%%%%%%%%%%%%%%%%%%%%%%%%%%%%%%%%%%%%%%%%%%%%%%%%%%%%%%%%%%%%%%%

Here, we discuss in detail the reaction-diffusion model
in Eqs.~(\ref{eq:BOdu}) and (\ref{eq:BOdvws}).
We start by considering the terms of the total ionic current $J$,
already discussed in Eq.~(\ref{eq:J}.
These ionic currents  are given by
\begin{subequations}
\begin{eqnarray}
J_{\rm fi} &=& \frac{v}{\tau_{\rm fi}}
          (u-\theta_v)(u_u -u) H_{\theta_v}(u) ,
\label{jfi}\\
J_{\rm si} &=& \frac{ws}{\tau_{\rm si}}H_{\theta_w}(u) ,
\label{jsi}\\
J_{\rm so} &=& -\frac{u}{T_o(u)}H^{\theta_w}(u)-
           \frac{1}{\tau_{\rm so1}+(\tau_{\rm so2} - \tau_{\rm so1})Q_{\rm so}(u)}
          H_{\theta_w}(u) \,,\label{jso}
\end{eqnarray}
\label{allJ}%
\end{subequations}
together with the stimulus current
\begin{eqnarray}
  J_{\rm stim} &=& j_{\rm stim} (H_0(t^{\ast})-H_0(t^{\ast}+T))\,,\label{jstim}
\end{eqnarray}
where $t^{\ast}=t\hbox{ mod}(1/f)$, $f$ being the frequency of
the stimulus signal, $0<T<1/f$ is the duration of the stimulus
and $j_{\rm stim}$ being its amplitude.
Figure \ref{fig06} illustrates each of the ionic current together
with the stimulus current and the normalized transmembrane
voltage. In our simulations we fix $j_{\rm stim}=-40$ pA and
$T=3.5$ ms, but similar results are obtained for other
stimulus conditions.
Function $H_x(z)$ is the Heaviside function, equal to $1$
for non-negative $z$ %-x$
and zero otherwise, and
$H^x(z)=1-H_x(z)$.

Equations (\ref{allJ}) (a)-(c) contain further the functions
\begin{subequations}
\begin{eqnarray}
T_x(u) &=& \tau_{x1}H^{\theta_x}(u) + \tau_{x2}H_{\theta_x}(u) \, ,\\
Q_x(u) &=& \frac{1}{2}\left (1+\tanh{(k_{x}(u-u_{x}))}\right ) \, ,
\end{eqnarray}
\label{otherfuncs}%
\end{subequations}
%for suitable choices of $x$,
where $u_u=1.0089$ and $u_{\rm so}=0.592093$ are reference values,
$\theta_v=0.3$ and $\theta_w=0.18171$ are threshold values of $u$
corresponding to the opening and closing of the ion channels,
$\tau_{\rm o1}=250.03$, $\tau_{\rm o2}=16.632$,
$\tau_{\rm so2}=6.5537$, 
and $k_{\rm so}=2.9748$.

As discussed in the main text, current $J_{\rm fi}$ is a fast inward
current mediated by sodium channels and controlled by the time scale
$\tau_{\rm fi}$, current $J_{\rm si}$ is a slow inward current mediated
by calcium channels and controlled by $\tau_{\rm si}$ and current
$J_{\rm so}$  is the slow outward current mediated by potassium
channels controlled by the time scale $\tau_{\rm so1}$.
Figures \ref{fig06} illustrates each ionic current as a function of
time, whereas in Fig.~\ref{fig07} we plot each current as a function
of the scaled potential $u$ and the three gating variables.
%The meaning of the currents is given in references
%\cite{Lenk/etal:2015,Richter:2017}.

Both Figs.~\ref{fig06} and \ref{fig07} may help understanding why
the set of the three time scales is suitable for characterizing
the full shape of one AP.
From Eq.~(\ref{jfi}) one sees that for voltages $u>\theta_v$
the fast inward current $J_{\rm fi}$ depends linearly on $v$
and quadraticaly on $u$. This current in time shows a very
narrow spike (Fig.~\ref{fig06}c) which results from a projection
over $u$ (Fig.~\ref{fig07}a). Thus, the quadratic dependence
in $u$ is not as dominant as the linear dependence on $v$
whose slope $1/\tau_{\rm fi}$ parameterizes the height of
the spike and consequently the amplitude of the AP.
This also explains why the amplitude $V_A$ depends more
strongly on $\tau_{\rm fi}$ than on the other time scales.

The slow inward current $J_{\rm si}$, Eq.~(\ref{jsi}),
is only relevant in the range $u>\theta_w$ (Fig.~\ref{fig06}d)
and, for that range, it depends linearly on both $w$ and $s$
gating variables (Fig.~\ref{fig07}b) with a slope given
by $1/\tau_{\rm si}$.

As for the slow outward current $J_{\rm so}$, Eq.~(\ref{jsi}),
it depends on $u$ exclusively.
It has two mutually exclusive regimes, one for $u<\theta_w$
and another for $u>\theta_w$.
As illustrated in Fig.~\ref{fig07}c, for $u<\theta_w$
the slow outward current evolves linearly to the transmembrane
current, with a slope given by a time scale, $\tau_{o1}$
or $\tau_{o2}$ depending if $u>\theta_o$ or $u<\theta_o$  respectively.
For $u>\theta_w$, the current $J_{\rm so}$ varies monotonically with
$u$, since it is a bounded step function of $u$ in the range $[0,1]$,
and consequently in this range of voltages $J_{\rm so}$ is governed
by one of the time scales, $\tau_{\rm so1}$ or $\tau_{\rm so2}$, which
we choose to be $\tau_{\rm so1}$.

These three time scales together with the three ionic currents
play also a role for defining the full model. As we will see next
the set of four equations can be reduced to only two
nonlinear equations, which include the dominant parts of
each ionic current, and consequently are tunnable through
their three time scales.

To see this we start by writing explicitly 
the three additional functions defining the evolution
of the gating variables in Eqs.~(\ref{eq:BOdu})
and (\ref{eq:BOdvws}):
\begin{subequations}
\begin{eqnarray}
E(u,v) &=&  -\frac{v-H^{\theta_v^-}(u)}{T^-_v(u)}H^{\theta_v}(u)-
             \frac{v}{\tau^+_v}H_{\theta_v}(u) \,, \label{eq:BOdv}\\[1ex]
F(u,w) &=&  -\frac{w-w_{\infty}}{\tau^{-}_ {w1}+
              (\tau^{-}_ {w2}-\tau^{-}_ {w1})Q^-_w(u)}H^{\theta_w}(u)-
             \frac{w}{\tau^+_w}H_{\theta_w}(u) \, , \label{eq:BOdw}\\[1ex]
G(u,s) &=&   \frac{1}{T_{\theta_w}(u)} \left ( Q_s(u)- s \right ),\label{eq:BOds}
\end{eqnarray}
\label{eq:BOFunctions}%
\end{subequations}
with
\begin{equation}
w_{\infty} = \left(1-\frac{u}{\tau_{w\infty}}\right) H^{\theta_o}(u)
          +w^*_{\infty} H_{\theta_o}(u) \, ,
\end{equation}
$T_x(u)$ and $Q_x(u)$ are given by Eqs.~(\ref{otherfuncs}), and
$\tau_{v1}^-=16.3$, $\tau_{v2}^-=1150$, $\tau_v^+=1.7026$, $\tau_{w1}^-=79.963$,
$\tau_{w2}^-=28.136$, $\tau_w^+=213.55$, $\tau_{w\infty}=0.2233$, 
$\tau_{\rm s1}=9.876$, and $\tau_{\rm s2}=4.2036$ are characteristic time
scales for the opening ($+$) and closing ($-$) of the ionic channels
(all in units of ms);
$w^{\ast}_{\infty}=0.902$, % is a limiting value $w$, 
$k_s=2.2268$, and $k_w^{-}=60.219$ are scaling parameters
and $u_s=0.81568$ and $u_w^{-}=9.991\times 10^{-3}$ are
the respective shape parameters for the hyperbolic
tangent in function $Q_x(u)$, and
$\theta_v^{-}=0.1007$ and $\theta_s=\theta_2$ are additional
threshold values for the opening and closing of the
ionic channels.

Figure \ref{fig08} shows the typical co-evolution of all variables
in the BOCF model, the scaled potential $u$ and the three
gating variables.

Next we will show that the BOCF model in
Eqs.~(\ref{eq:BOdu}) and (\ref{eq:BOdvws})
can be treated in a semi-analytically way
for $D=0$ (single cell case), by properly introducing approximations
of the equations in the $u$-regions defined through the Heaviside
functions (cf.~Fig.~\ref{fig08}a), namely
\begin{itemize}
\item Region I where $\theta_v < u < 1$,
\item Region II where $\theta_w < u < \theta_v = 0.3$,
\item Region III where $\theta_v^- < u < \theta_w = 0.18171$,
\item Region IV where $\theta_o < u < \theta_v^- = 0.1007$ and
\item Region V where $0 < u < \theta_o = 0.015473$.
\end{itemize}
Substituting the limiting values above in the currents defined in
Eqs.~(\ref{allJ}) and in the functions defined in
Eqs.~(\ref{eq:BOFunctions}) yields a system of four differential
equations for each region.

At the beginning of each AP, the stimulus current
$J_{\rm stim}$ is applied bringing $u$ to its maximum value,
$u\sim 1$, i.e.~in region I. From there on, the systems
evolves according to Eqs.~(\ref{eq:BOdu}) and (\ref{eq:BOdvws})
till the next stimulus (see inset of Fig.~\ref{fig08}a).

In region I the dynamical equations read
\begin{subequations}
\begin{eqnarray}
  \frac{du}{dt} &=& -\frac{1}{\tau_{\rm so1}+(\tau_{\rm so2}-\tau_{\rm so1})Q_{\rm so}(u)}
                    + \frac{ws}{\tau_{\rm si}}+                
                     \frac{v}{\tau_{\rm fi}}(u-\theta_v)(u_u-u) \, ,\label{uR1}\\
  \frac{ds}{dt} &=&  \frac{Q_s(u)-s}{\tau_{s2}} \, ,\label{sR1}
\end{eqnarray}
\label{R1}%
\end{subequations}
where $v=V_{u=1}\exp{\left ( -t/\tau_v^+ \right )}$
and $w=W_{u=1}\exp{\left ( -t/\tau_w^+ \right )}$
decay exponentially and independently of all other
variables.
In other words, the evolution of the four dimensional systems
reduces to a nonlinear and non-autonomous two-dimensional system
of coupled variables, $u$ and $s$.

As will become clear below,
this dynamical system (\ref{R1}) is the only part of the model
equations that cannot be solved in closed analytical form, while
the behavior in other regions becomes analytically tractable
after proper approximations.
Notice that the Eq.~(\ref{uR1}), defining the time evolution
of the normalized action potential $u$, is composed by three
contributions, each on corresponding to one of the three ionic
currents and being parameterized by one of the three time scales.
See Eq.~(\ref{allJ}) and the discussion above.

In region II, both variables $w$ and $s$ are governed by the same
equations as in region I,
while the potential variable $u$ has no longer the quadratic term
(see Eq.~(\ref{uR1})). As for the variable $v$, it decays 
exponentially with a different constant $\tau_{v2}^-$.
Since in region I the decay of $v$ is strong enough for bringing
$v$ close to zero, one can approximate $v=0$ in region II and
consequently the time evolution of $u$ is approximated by Eq.~(\ref{uR1}).

In region III, $u$ and $v$ decay exponentially as
$u=\theta_w\exp{\left ( -t/\tau_{\rm o2}\right )}$
and $v=V_{u=\theta_w}\exp{\left ( -t/\tau_{v2}^- \right )}$, and
$w$ and $s$ are coupled to each other according to
the two-dimensional system
\begin{subequations}
\begin{eqnarray}
  \frac{dw}{dt} &=& -\frac{w-w_{\infty}^{\ast}}{\tau_{w1}^-+(\tau_{w2}^--\tau_{w1}^-)Q_{w}^-(u)} \, ,\label{wR3}\\
  & & \cr
   \frac{ds}{dt} &=&  \frac{Q_s(u)-s}{\tau_{s1}} \, .\label{sR3}
\end{eqnarray}
\label{R3}%
\end{subequations}
For this range of $u$ values, $Q_w^-(u)\sim 1$ and
$Q_s(u)$ is almost constant.
Therefore, we can set
$Q_s(u)\sim \langle Q_s(u)\rangle_{\theta_v^-<u<\theta_w}\equiv 
\langle Q\rangle = 0.0475$ and
consequently $w$ and $s$ are approximately given by
\begin{subequations}
\begin{eqnarray}
  w(t) &=& w_{\infty}^{\ast} + \left ( W_{u=\theta_w}-w_{\infty}^{\ast} \right ) 
           \exp{\left ( -t/\tau_{w2}^- \right )} \, ,\label{wR3sol}\\
  & & \cr
  s(t) &=& \langle Q\rangle + \left ( S_{u=\theta_w}-\langle Q\rangle \right ) 
           \exp{\left ( -t/\tau_{s1} \right )} \, .\label{sR3sol}
\end{eqnarray}
\label{R3sol}%
\end{subequations}

In region IV,
$u$ and $v$ decay exponentially as
$u=\theta_v^-\exp{\left ( -t/\tau_{\rm o2}\right )}$
and 
\begin{equation}
v=1+\left ( V_{u=\theta_v^-}-1
         \right )
         \exp{\left ( -t/\tau_{v1}^- \right )},
\end{equation}
respectively.
The gate variable $s$ follows the same approximation
as in Region III, Eq.~(\ref{sR3sol}).
The variable $w$ follows the same Eq.~(\ref{wR3}),
but now with a different approximation, namely
\begin{equation}
R(t)\equiv \frac{1}{\tau_{w1}^-+(\tau_{w2}^--\tau_{w1}^-)Q_w^-(u)} 
\sim 1 - \alpha \hbox{\large{e}}^{-2k_w^-(u-u_w^-)} \, ,
\end{equation}
with
\begin{equation}
  \alpha = \frac{\tau_{w1}^-+(\tau_{w2}^--\tau_{w1}^-)Q_w^-(\theta_o)-1}{\tau_{w1}^-+(\tau_{w2}^--\tau_{w1}^-)Q_w^-(\theta_o)} 
           \hbox{\large{e}}^{2k_w^-(\theta_o-u_w^-)} \, .
\end{equation}
%yielding
%\begin{equation}
%w(t) = w_{\infty}^{\ast} + 
%     K \exp{\left ( -\int_t R(t^{\prime})dt^{\prime} \right )} \, .
%\label{wR4sol}
%\end{equation}
Since in this region, the values of $u$ are small and the time-window is
also small, the exponential decay of $u$ can be linearized,
$u\sim \theta_v^- (1-t/\tau_{o2})$, which gives
\begin{equation}
R(t) \sim 1 - \alpha \Gamma_1 \hbox{\large{e}}^{\Gamma_2t} \sim 1 - \alpha\Gamma_1 -\alpha\Gamma_1\Gamma_2 t\, 
\end{equation}
with
\begin{subequations}
\begin{eqnarray}
  \Gamma_1 &=& \hbox{\large{e}}^{-2k_w^-\theta_v^-(1-u_w^-)} \, ,\\
  \Gamma_2 &=& 2\frac{k_w^-\theta_v^-}{\tau_{o2}} \, .
\end{eqnarray}
\end{subequations}
This approximation yields for the evolution of $w$ in this
region
\begin{equation}
w(t) = w_{\infty}^{\ast} + 
      (W_{u=\theta_v^-}-w_{\infty}^{\ast})
      \exp{\left ( -(1-\alpha\Gamma_1)t+\frac{\alpha\Gamma_1\Gamma_2}{2}t^2 \right )} \, .
\label{wR4sol2}
\end{equation}
%where $Ei(\cdot)$ is the exponential integral
%\begin{equation}
%Ei(-x) = \int_{x}^{\infty} \frac{\hbox{\large{e}}^{-y}}{y}dy \, ,
%\end{equation}
%that can be solved numerically, and
%\begin{subequations}
%\begin{eqnarray}
%\gamma_1 &=& \frac{\tau_{\rm o2}}{k_w^-} \left (
%                                   1-\alpha \hbox{\large{e}}^{2k_w^-u_w^-}
%                                   \right ) , \label{gamma1}\\
%   & & \cr
%\gamma_2 &=& -\frac{\tau_{\rm o2}}{k_w^-} \alpha , \label{gamma2}\\
%   & & \cr
%K &=& \frac{w(\theta_v^-)-w_{\infty}^{\ast}}
%           {(\theta_v^-)^{\gamma_1}
%            \exp{\left ( \gamma_2 Ei(-2k_w^-\theta_v^-) 
%                 \right )}} \label{K} .
%\end{eqnarray}
%\end{subequations}

Finally, in region V, variables $u$, $v$ and $s$ follow the
same solution as in region IV but for different constants,
namely $u$ decays exponentially with decay time
$\tau_{\rm o1}$ instead of $\tau_{\rm o2}$,  
and 
$Q_s(u)\sim \langle Q_s(u)\rangle_{0<u<\theta_o} = 0.02665$.
The remaining gate variable $w$ is approximated
by observing (see Fig.~\ref{fig08}a) that in this range
$u\sim 0$ and $Q_s(u)$ can be set to a constant
$Q_s(0)$, yielding
\begin{equation}
w(t) = 1 + \left ( W_{u=\theta_o} -1
           \right ) \exp{\left ( 
                         -t/T
                         \right )} 
\label{wR5sol}
\end{equation}
with
\begin{equation}
T = \tau_{w1}^-+(\tau_{w2}^--\tau_{w1}^-)Q_w^-(0) .
\end{equation}

Altogether, we arrive to the conclusion that the problem of
solving the single-cell dynamics of the BOCF model (\ref{eq:BOdu})
and (\ref{eq:BOdvws})
can be reduced to the two-dimensional non-linear system
in Eqs.~(\ref{R1}), which involves the three time scales
controlling each ionic current considered in the BOCF model.
%Notice that the analysis of possible stationary solutions of this
%model is not meaningful in this scope since, due to the external
%current $J_{\rm stim}$ the systems' evolution is non-stationary.

%%%%%%%%%
\section*{Acknowledgments}

The authors thank C.~Lenk and G.~Seemann for helpful discussions and
the Deutsche Forschungsgemeinschaft for
financial support (Grant no.\ MA1636/8-1).

%\nolinenumbers

%%%%%%%%%%%%%%%%%%%%%%%%%%%%%%%%%%%%%%%%%%%%%%%%%%%%%%%%%%%%%%%%%%%
%%%%%%%%%
%\bibliographystyle{plos2015}
%\bibliographystyle{apsrev4-1}
%\bibliography{literature}

%%%%%%%%%%%%%%%%%%%%%%%%%%%%%%%%%%%%%%%%%%%%%%%%%%%%%%%%%%%%%%%%%%
%%%%%%%%%%%%%%%%%%%%%%%%%%%%%%%%%%%%%%%%%%%%%%%%%%%%%%%%%%%%%%%%%%
\end{document}